# Microtesla MRI of the human brain combined with MEG

Vadim S. Zotev [*], Andrei N. Matlashov, Petr L. Volegov, Igor M. Savukov,
Michelle A. Espy, John C. Mosher, John J. Gomez, Robert H. Kraus Jr.

*Los Alamos National Laboratory, Applied Modern Physics Group, MS D454, Los Alamos, NM 87545, USA*

**Abstract**

One of the challenges in functional brain imaging is integration of complementary imaging modalities, such as magnetoencephalography (MEG) and functional magnetic resonance imaging (fMRI). MEG, which uses highly sensitive superconducting quantum interference devices (SQUIDs) to directly measure magnetic fields of neuronal currents, cannot be combined with conventional high-field MRI in a single instrument. Indirect matching of MEG and MRI data leads to significant co-registration errors. A recently proposed imaging method – SQUID-based microtesla MRI – can be naturally combined with MEG in the same system to directly provide structural maps for MEG-localized sources. It enables easy and accurate integration of MEG and MRI/fMRI, because microtesla MR images can be precisely matched to structural images provided by high-field MRI and other techniques. Here we report the first images of the human brain by microtesla MRI, together with auditory MEG (functional) data, recorded using the same seven-channel SQUID system during the same imaging session. The images were acquired at 46 microtesla measurement field with pre-polarization at 30 mT. We also estimated transverse relaxation times for different tissues at microtesla fields. Our results demonstrate feasibility and potential of human brain imaging by microtesla MRI. They also show that two new types of imaging equipment – low-cost systems for anatomical MRI of the human brain at microtesla fields, and more advanced instruments for combined functional (MEG) and structural (microtesla MRI) brain imaging – are practical.

## 1. Introduction

Detailed understanding of human brain function requires the ability to perform noninvasive imaging of brain activity with both high temporal and high spatial resolution. No single imaging modality can satisfy both requirements at present. Magnetoencephalography (MEG) [1] and electroencephalography (EEG) measure the direct consequences of neuronal activity with millisecond temporal resolution, but their source localization accuracy is limited due to the ill-posed nature of the electromagnetic inverse problem. Moreover, these methods cannot image brain structure, which is usually obtained by a separately performed magnetic resonance imaging (MRI) [2]. Functional MRI (fMRI) [3-5] can provide high spatial resolution, but its temporal resolution is limited by the natural slowness of the hemodynamic response. Moreover, the relationship between such response and underlying neuronal activity is not yet fully understood [6].

Integration of complementary imaging modalities [7], e.g., the combination of fMRI with MEG [8] or EEG [9], is commonly viewed as an approach to realize high-resolution spatiotemporal imaging of brain function. Such integration goes beyond the simple addition of capabilities, because fMRI data can be used to bias solution of the inverse problem, and thus improve accuracy of MEG/EEG localization [8]. Numerous comparative studies of fMRI and MEG, e.g. [10,11], have suggested that combination of these methods should be used for more reliable clinical diagnosis.

However, high magnetic fields and intense rf pulses employed in conventional MRI make its direct combination with other techniques extremely difficult. While simultaneous acquisition of EEG and fMRI signals is technically challenging [9], combination of MEG, which uses superconducting quantum interference devices (SQUIDs) [12], and MRI in a single instrument is practically impossible. MEG and MRI (or fMRI) data, acquired by two completely different systems, can only be matched indirectly by means of an elaborate, time-consuming, and error-prone co-registration procedure [13]. Typical MEG/MRI co-registration errors of the order of 5-10 mm [13] exceed average MEG source localization errors [14] and make MEG less efficient as a clinical evaluation tool.

SQUID-based magnetic resonance imaging at microtesla fields, also referred to as ultralow-field (ULF) MRI, is a promising new imaging method, introduced by the UC Berkeley researchers [15,16]. It uses magnetic sensors of the same type – SQUIDs with untuned input circuits – as those used for MEG [15]. ULF MRI can be naturally combined with MEG in the same system to directly provide anatomical maps for MEG-localized neural sources [15,16]. It has been demonstrated by our group that MEG and ULF NMR signals can even be acquired simultaneously using the same SQUID [17]. We have also developed multichannel SQUID instrumentation for both MEG and ULF MRI [18-20]. However, no implementation of the combined MEG and ULF MRI of the human brain has been reported until now.

---
[*] Corresponding author. Fax: +1 505 665 4507.
E-mail address: vzotev@lanl.gov (V.S. Zotev).

In addition to the MEG-style untuned SQUID detection of MRI signals, the ULF MRI method relies on the pre-polarization technique [21,22] to increase sample magnetization prior to each imaging step performed at a microtesla-range measurement field [15]. In contrast to conventional MRI, relative homogeneity of the measurement field is not crucial in ULF MRI, because microtesla-range magnetic fields of even modest relative homogeneity are highly homogeneous on the absolute scale [23].

Microtesla MRI holds three important promises for medical imaging in general, and neuroimaging in particular. First, imaging at ULF can be performed using simple, inexpensive, and portable coil systems of open geometry [19,24], that do not subject a patient to high magnetic fields of conventional MRI and allow imaging in the presence of metal [25]. Such systems can make MRI more affordable and better suited for operating rooms and field hospitals. Second, $T_1$-weighted contrast usually improves at low magnetic fields [26], and strong magnetic relaxation dispersion exhibited by tissues in the µT – mT field range [27] can be used to selectively enhance this contrast. This may allow more efficient identification of various medical conditions that affect $T_1$, such as brain tumors [28], without the use of potentially toxic [29] gadolinium based contrast agents. Third, as mentioned above, microtesla MRI can be combined with MEG and other SQUID-based techniques for biomagnetic measurements [30]. This allows development of new medical instruments, such as multichannel SQUID systems for both functional (MEG) and structural (ULF MRI) imaging [18] of the human brain. In a parallel effort, existing whole-head MEG systems can be modified [31] to include ULF MRI capability. Microtesla MRI might also enable direct tomographic imaging of neural currents [32].

In this paper, we report the first images of the human brain acquired by microtesla MRI. We also present auditory MEG data recorded using the same multichannel SQUID system during the same imaging session. Such instruments for combined MEG/ULF-MRI can substantially benefit neuroimaging. No co-registration of MEG and ULF MRI data is required after spatial sensitivities of MEG sensors are mapped by ULF MRI during an initial uniform-phantom calibration. Because ULF images can be precisely matched to structural images provided by other imaging modalities, ULF MRI can enable seamless integration of MEG and EEG, on the one hand, with high-spatial-resolution MRI and fMRI, on the other. Moreover, the potential of ULF MRI as an integration tool is not limited to these methods. Various medical techniques are easier to combine in the same instrument with ULF MRI, than with conventional MRI, as discussed below. Matching ULF and conventional MR images makes it possible to provide high-quality structural maps for almost any medical procedure that might benefit from simultaneous anatomical imaging. ULF MRI might also help to integrate various medical techniques with imaging methods other than MRI, e.g. X-ray computed tomography (CT) and positron emission tomography (PET).

## 2. Methods

The brain imaging results, reported in this paper, were obtained using the experimental system [18-20] and measurement procedure, depicted schematically in Fig. 1. The system includes seven second-order SQUID gradiometers with magnetic field resolution of 1.2 – 2.8 fT/√Hz, installed inside a flat-bottom liquid helium cryostat in a pattern shown in Fig. 1A. The gradiometers have 37 mm coil diameter, 60 mm baseline, and 45 mm center-to-center spacing for the neighboring coils. The cryostat is mounted inside an open-type coil system, Fig. 1B, that generates magnetic fields and gradients for ULF MRI according to the sequence in Fig. 1C. Each of the five sets of coils in Fig. 1B is symmetric with respect to the center of the system, and the largest ($B_m$) coils are 120 cm in diameter. The system is operated inside a magnetically shielded room, which makes it possible to perform MEG measurements in addition to ULF MRI. Technical details of our instrumentation have been previously reported [19].

The ULF images were acquired at the measurement field $B_m$=46 µT, which is similar in strength to the Earth's magnetic field and corresponds to the proton Larmor frequency of about 2 kHz (for protons, $\gamma/2\pi$=42.6 Hz/µT). A stronger pre-polarizing field $B_p$=30 mT was applied for 1 s prior to each imaging step, and was removed before the application of $B_m$.

The 3D imaging sequence used in our ULF MRI experiments (Fig. 1C) does not employ any rf pulses. Instead, simple manipulations with the measurement field $B_m$ are performed. Spin precession is induced by application of $B_m$ perpendicular to the original direction of $B_p$ [19], and echo is generated by $B_m$ reversal [24]. This method greatly simplifies the ULF MRI instrumentation. The encoding scheme is based on the 3D Fourier protocol with frequency encoding gradient $G_x=dB_z/dx$ and two phase encoding gradients, $G_z=dB_z/dz$ and $G_y=dB_z/dy$ [19,20]. The following imaging parameters were used in the present work: $B_p$=30 mT, $B_m$=±46 µT, $G_x$=±140 µT/m, $|G_z|\leq140$ µT/m, 61 encoding steps, $|G_y| \leq 70$ µT/m, 11 steps, $t_p$=1 s, $t_g$=28 ms, and $t_a$=56 ms. This imaging sequence provided 3 mm × 3 mm × 6 mm spatial resolution. To study transverse relaxation properties of brain tissues, we implemented a modification of the commonly used multiple-echo technique [33]. Each echo is induced by simultaneous reversal of the measurement field $B_m$ and the readout gradient $G_x$ (Fig. 1C) to compensate for spatial inhomogeneities of both. This approach allows us to measure $T_2$, rather than $T_2^*$ relaxation times. Unlike the 180º rf pulse in typical spin echo sequences, the reversal of $B_m$ does not compensate for residual field inhomogeneities. The residual fields, however, are negligible inside our magnetically shielded room. The echo time $TE$ is measured from the moment $B_m$ is first applied. Four echoes with $TE$=63, 142, 205, and 283 ms, respectively, were acquired at each imaging step.

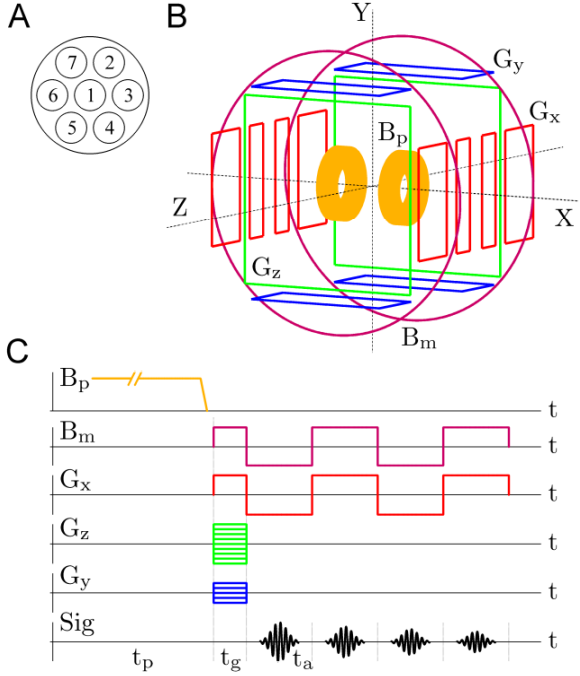

**Fig. 1.** Experimental set-up and procedure for multiecho 3D ULF MRI. (A) Positions of the seven SQUID gradiometers inside the liquid He cryostat. (B) Schematic layout of the coil system for 3D Fourier imaging with pre-polarization. (C) Multiple-echo 3D imaging sequence.

A complete scan of the phase space in our ULF MRI experiments included 61×11 encoding steps and required 15 minutes. To improve image quality, six consecutive scans were performed, and the resulting images were averaged. The total imaging time was about 90 minutes in each experiment, with 75% of this time taken up by pre-polarization. We discuss possible approaches to improve the system SNR and reduce imaging time in Section 4.

All experiments involving human subjects were approved by the Los Alamos Institutional Review Board, and informed consent was obtained from the subject involved. The human subject was positioned comfortably inside the system with the head between the $B_p$ coils under the bottom of the cryostat. Fig. 2 shows orientation of the head with respect to the seven pick-up coils during each imaging session. In the first ULF MRI experiment, the head was positioned to ensure coverage of important anatomical features by the system channels (Fig. 2A). In the second experiment, the cryostat was turned 90 degrees, and the forehead area was imaged as illustrated in Fig. 2B.

The MEG measurements were performed immediately after the ULF MRI of the right side of the head, while the human subject remained inside the system. The head was repositioned slightly as shown in Fig. 2C to increase coverage of the auditory cortex. Such head repositioning would not be necessary with whole-head SQUID arrays typically used in MEG. The auditory stimulus consisted of a 50 ms long 1 kHz tone pulse with 500 ms pre-stimulus interval. The MEG sequence was repeated 200 times, and auditory evoked responses measured by each channel were digitally filtered and averaged. All the magnetic fields and gradients used in ULF MRI were turned off during the MEG experiment. It would also be possible to alternate MEG measurements and ULF MRI imaging steps.

The high-field 3D image of the same human subject's head was acquired by conventional MRI at 1.5 T using spin-echo sequence with $TE$=64 ms and $TR$=9000 ms. The image, originally with 1 mm isotropic resolution, was subjected to rotation and summation over depth within 6 mm-thick layers to approximately match each of the two 3D ULF MR images below.

## 3. Results

The first images of the human head by microtesla MRI are exhibited in Figs. 3 and 4. The right side of the head (Fig. 2A) and the forehead area (Fig. 2B) were imaged as described above. Each ULF image in Figs. 3 and 4 is a composite image computed as a square root of the sum of squares of images from the seven SQUID channels of our system. Each image was also subjected to fine-mesh interpolation and correction of concomitant gradient artifacts [20]. Only four horizontal image layers (out of 11 simultaneously acquired) at depths *D*=18 to 36 mm are shown in Fig. 3. The ULF images in this figure are accompanied by images of the same human subject's head provided by conventional MRI as explained above. Comparison of the ULF and high-field MR images demonstrates that ULF MRI can be successfully used for structural imaging of the human head.

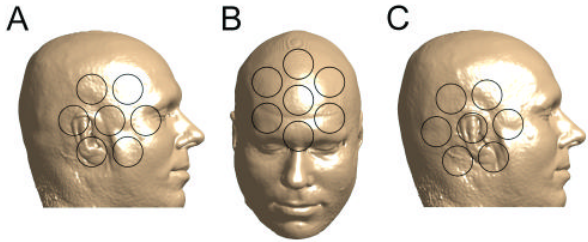

**Fig. 2.** Positions of the human subject's head with respect to the seven SQUID channels during (A) microtesla MRI of the right side of the head; (B) microtesla MRI of the forehead area; (C) auditory MEG measurements.

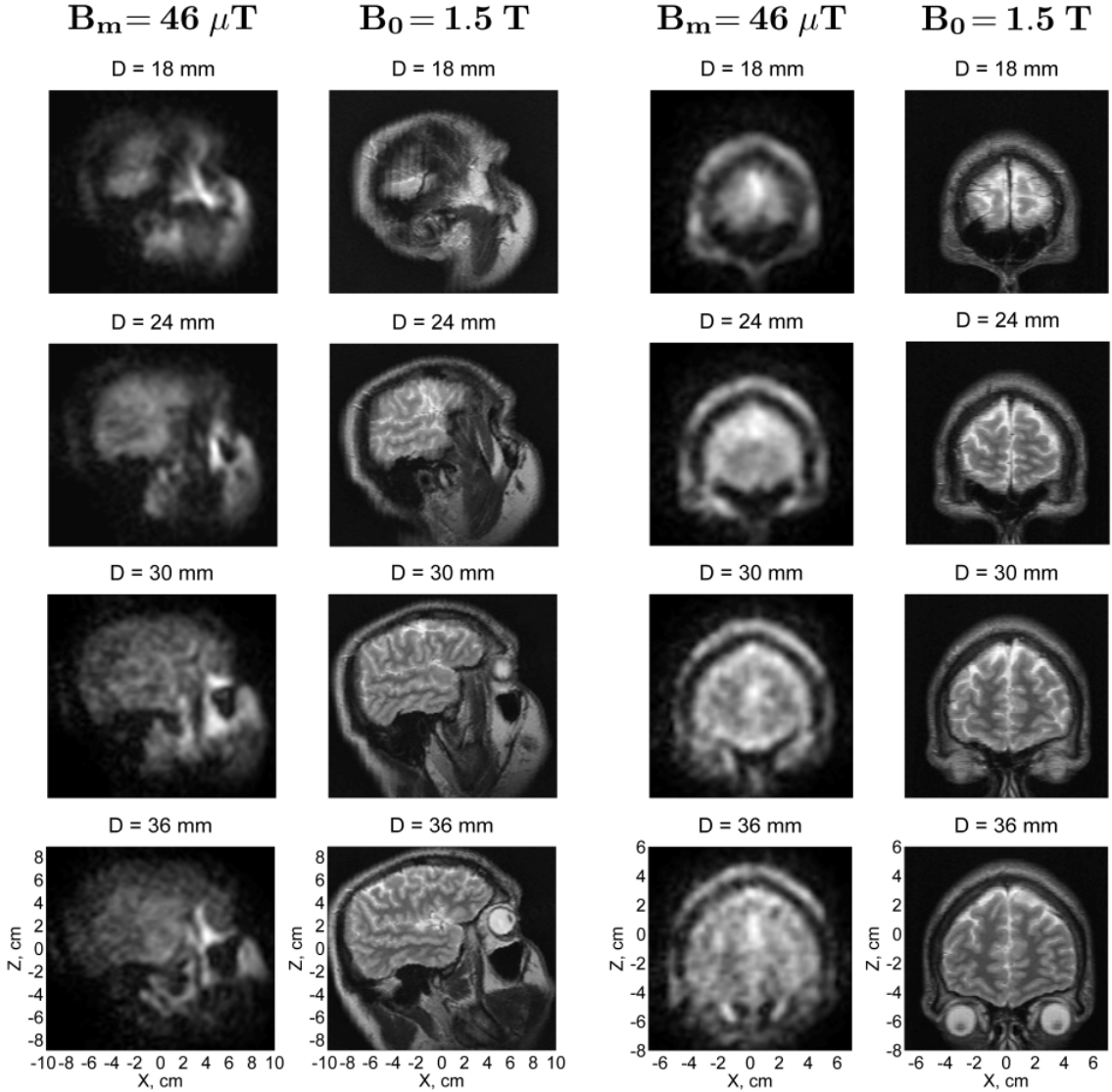

**Fig. 3.** Microtesla MRI of the human head compared to conventional MRI. The 3D ULF MR images of the head side and the forehead area were acquired at 46 µT measurement field. Each image in the figure represents a 6 mm-thick layer of the head. *D* is the depth of the central plane of a given layer with respect to the bottom of the cryostat. The in-plane resolution is 3 mm × 3 mm. The high-field 3D image of the same subject's head was acquired by conventional MRI at 1.5 T. Each high-field image in the figure corresponds to the same layer of the head as the ULF image on its left.

The ULF images in Fig. 3 correspond to the first echo with *TE*=63 ms. Brain tissues (gray and white matter) in these images have approximately the same brightness as cerebrospinal fluid (CSF), which can be explained as follows. Because the pre-polarization time of 1 s is longer than $T_1$ values in the brain (for white matter, $T_1$=200 ms was reported at 20 mT field [34]), but shorter than $T_1$ of CSF, the initial polarization of brain tissues is higher than that of CSF. However, $T_2$ for CSF is longer at 46 µT than $T_2$ for gray and white matter, as shown below. Thus, the brain tissues in our experiments should look brighter at shorter echo times, while the CSF should be brighter at longer *TE*s. The latter tendency is observed in Fig. 4, which shows images corresponding to four consecutive echoes.

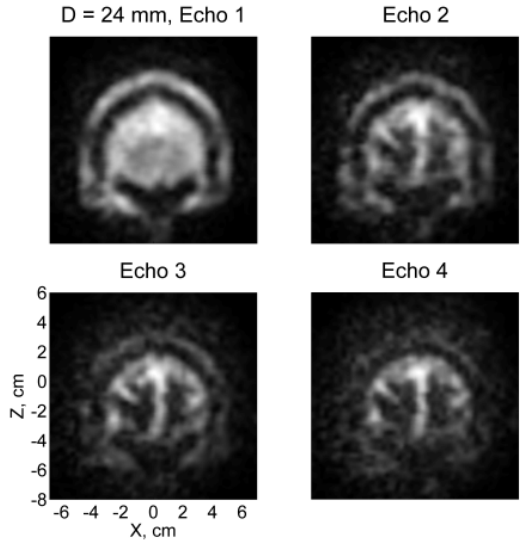

**Fig. 4.** $T_2$–weighted contrast in microtesla MRI of the human head. The four images of the same 6 mm-thick layer correspond to four consecutive echoes with $TE$=63, 142, 205, and 283 ms.

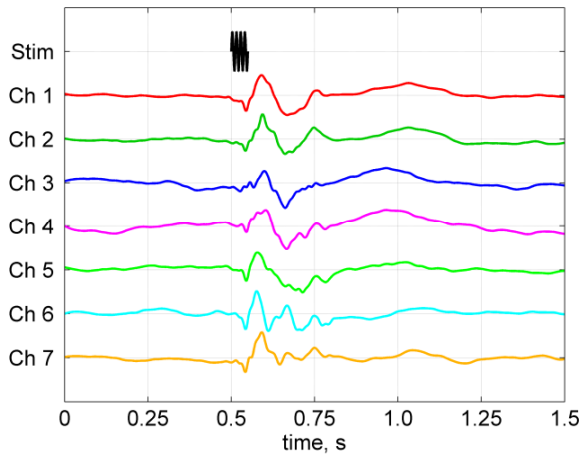

**Fig. 5.** Auditory MEG recordings with the same system. The auditory evoked response curves have peak-to-peak magnitudes of 237, 132, 81, 78, 95, 149, and 239 fT for channels 1 through 7, respectively. They are normalized by one in the figure to emphasize their time dependence.

According to Fig. 4, the $T_2$-weighted contrast at 46 µT field between brain tissues and CSF improves visibly with echo time.

Using multiple-echo data, we estimated $T_2$ values for different human tissues for the first time at ULF. Identification of tissues was based on a detailed comparison of ULF images with high-field MR images of the same subject's head (Fig. 3), together with analysis of their long-time relaxation (Fig. 4). Image intensities for 10-20 pixels corresponding to a certain tissue type were then averaged for each echo, and a single exponential function was fitted to the resulting data. The error bars below represent standard deviations of the respective fits. For gray and white matter, $T_2$ values at 46 µT were found to be 106±11 ms and 79±11 ms, respectively. The relaxation is slower for CSF, with $T_2$=355±15 ms. Other $T_2$ values easily determined from our data are 120±7 ms for scalp, 102±5 ms for maxillary sinuses, 108±2 ms for soft tissues around eyes, and 667±23 ms for vitreous bodies of the eyes. In each case, the same analysis was also applied to images of a large uniform water phantom, and relaxation times longer than 1400 ms were invariably obtained. This means that $T_2$ values determined in this work are sufficiently reliable, and not shortened significantly due to field inhomogeneities or sequence imperfections.

Based on these relaxation measurements, we conclude that $T_2$ values for gray and white matter at 46 µT are similar to those measured in conventional MRI. At 3T field, for example, $T_2$ is equal to 110 ms and 80 ms for gray and white matter, respectively [35]. It should be noted that earlier MRI studies of magnetic relaxation in the brain tended to yield shorter $T_2$ values [36]. Relaxation properties of CSF are similar to those of water, with $T_2$=1.76 s reported at 0.15 T [37]. However, CSF $T_2$ times shorter than 1 s have appeared quite often in literature. Our $T_2$ value of 355±15 ms for CSF is relatively short, which can be attributed to partial volume effects due to the relatively large voxel size in our experiments. In contrast, the $T_2$ time of 667±23 ms for the vitreous body at 46 µT is longer than the average value of $T_2$=390 ms reported at 1.5 T [38]. The reason for this difference is unclear. To conclude, our results generally agree with earlier observations that $T_2$ does not exhibit a strong dependence on magnetic field strength [28,35]. This does not mean, however, that $T_2$-weighted contrast between two specific tissues would not change, to some degree, with magnetic field. Further and more extensive studies of transverse relaxation at ULF are needed because of the important medical role played by $T_2$ contrast. Moreover, because $T_1$ and $T_2$ are expected to converge at low frequencies in the motional-narrowing regime [27], $T_2$ values should approximate longitudinal relaxation times $T_1$ at microtesla fields. Systematic in vivo studies of $T_1$ relaxation dispersion in the human brain and other organs in the µT – mT field range are also very important, as explained in the introduction.

Results of the auditory MEG measurements with the same seven-channel SQUID system are shown in Fig. 5. Each auditory evoked response curve in Fig. 5 exhibits a series of peaks characteristic of auditory MEG [1]. This result demonstrates that our system can be used for both ULF MRI of the human brain and magnetoencephalography. The magnitudes of MEG signals in Fig. 5 suggest that the equivalent current dipole was located in the general vicinity of channels 1 and 7 (the numbering scheme in Fig. 2C is the same as in Fig. 1A). It should be noted, however, that the gradiometer pick-up coils in our system are larger (37 mm diameter) than in commercial MEG instruments, so the source localization accuracy is inevitably lower. Sensor arrays in whole-head MEG/ULF-MRI systems will have to be optimized to ensure high MEG localization accuracy, on the one hand, and good ULF MRI depth sensitivity and parallel imaging performance, on the other.

## 4. Discussion

In the previous section, we reported the following results: 1) the anatomical images of the human head acquired by the low-cost ULF MRI system; 2) the analysis of $T_2$ relaxation in the human head at ULF; 3) the combination of MEG and brain ULF MRI capabilities in one instrument. Each of these results is obtained for the first time. Further studies in each of these directions would be very beneficial.

The combination of MEG and ULF MRI appears to have the most immediate practical significance. Even though the imaging resolution at ULF reported here (Fig. 3) is not high, it allows 3D matching of ULF images (and any related MEG data) to high-field MR images of the same head with better accuracy than that of the traditional MEG/MRI co-registration [13]. Therefore, the main advantage of combined MEG and ULF MRI can be demonstrated at present, while ULF MRI is still at an early stage of development. Our next research goal is to investigate localization of primary somatosensory, motor, auditory, and visual functional areas by MEG, and acquire structural maps of the same areas by ULF MRI. We will then perform direct 3D superposition of MEG and ULF MRI data, and determine the accuracy of matching ULF and conventional MR images that can be achieved in practice.

The fact that ULF MRI can be combined in the same instrument with such a demanding technique as MEG suggests that it can also be combined with a variety of other medical techniques. This would be particularly beneficial in the case of surgical and interventional procedures, which are increasingly often performed under the guidance of conventional MRI [39,40]. Examples of such procedures include neurosurgery, biopsy, endoscopy, targeted drug delivery, intravascular therapy, various ablative therapies, etc. To ensure an easy access to the patient, partially open MRI scanners with 0.1 T – 0.6 T magnetic fields are widely used for intraoperative and interventional MRI [39,40]. ULF MRI offers unique advantages in this respect. A system of the measurement field and gradient coils for ULF MRI can be large enough to comfortably accommodate inside both the patient and the physician, and open enough to allow access from any direction. Positions and orientations of both the cryostat and the pre-polarization magnet can be adjustable for maximum operating convenience and imaging efficiency. Because image distortions around metal pieces and rf heating are greatly reduced at low fields [22,25], many common medical instruments can be safely used inside a ULF MRI system. As argued in the introduction, ULF images can be substituted with higher-resolution conventional MR images of the same area using the simple image matching (provided that the underlying anatomy does not change during the procedure).

As mentioned in Section 2, each ULF image reported in this work is an average of six single-scan images. We expect to be able to improve our system's SNR sufficiently to acquire good-quality images without averaging. Moreover, one can take advantage of distinct spatial sensitivities of different coils in a sensor array to achieve imaging acceleration. Our seven-channel system allows imaging with the acceleration factor $R$=3 [20], based on 1D phase space undersampling and SENSE image reconstruction [41]. Thus, the single-scan imaging time can be reduced from 15 to 5 minutes. Because the image SNR scales as the square root of the total acquisition time, the reduction in total imaging time from the present 90 minutes to 5 minutes will require an improvement in intrinsic system SNR by a factor of 4. Such an improvement is well within our present technological capabilities.

The SNR of ULF MRI instruments can be increased through the use of stronger pre-polarizing fields and reduction in system noise. Our practical experience with state-of-the-art SQUID technology suggests that one can achieve magnetic field resolution of a fraction of 1 fT/√Hz by using larger pick-up coils. Similarly, a special cryostat design makes it possible to lower Johnson noise of the thermal shield to sub-femtotesla levels. Another source of Johnson noise in SQUID instruments is the rf shield, which usually consists of gold- or aluminum-plated mylar placed inside or outside the cryostat. It is not clear at present whether its noise can be reduced substantially without impairing its rf screening properties. We can already demonstrate sub-femtotesla field resolution with our second-generation system, but it remains to be seen how far below 1 fT/√Hz we can actually go. The highest pre-polarization field presently achieved is 0.4 T, as reported by the Stanford group [22]. It was generated by a compact resistive magnet with 9 cm diameter bore, and used to perform pre-polarized imaging around metal orthopedic implants [22]. This result suggests that it is possible to design pre-polarizing coils for human head imaging that produce magnetic fields of 0.1 T and higher. Finally, it is important to note that ULF MRI technology is only beginning to develop, and further improvements to various instrumentation components should significantly enhance the overall imaging performance.

ULF MRI will also greatly benefit from parallel imaging by multichannel SQUID systems. As we argued before [20], parallel imaging is easier to implement in ULF MRI than in conventional high-field MRI, because the untuned SQUID detection of ULF MRI signals makes inductive decoupling of pick-up coils unnecessary. Moreover, system noise is essentially sample-independent at ULF, and noise correlations among SQUID channels can be very low, provided that the cryostat is properly designed [20]. The biomagnetism community has large experience in building commercial MEG instruments with hundreds of SQUID channels. If MRI signal from a given voxel is received simultaneously by $N$ coils with similar sensitivities at that point and uncorrelated noise, the SNR improvement scales as $\sqrt{N}$ (see, e.g., Eq. (5) in [20]). If an array of $N$ coils is used for accelerated imaging, the maximum acceleration factor is $R=N$ in SENSE method [41], though practical values of $R$ are usually lower. Based on these considerations, we expect ULF MRI systems with 150-300 SQUID channels to allow much faster imaging than can be presently achieved with one- or seven-channel instruments. One should keep in mind, however, that parallel imaging performance strongly depends on array geometry, and each array configuration has to be simulated and optimized individually.

We would also like to mention that the advantages of microtesla MRI, including its combination with MEG, can also be realized if optical atomic magnetometers, that do not require cryogenic cooling, are used to measure magnetic signals [42,43]. Further improvements in such devices and studies of their potential as an alternative to SQUIDs for these applications would be most interesting.

## 5. Conclusion

The results reported in this paper, together with the other results in this field, suggest that SQUID-based microtesla MRI is becoming a new brain imaging modality with its own unique opportunities and challenges. Its further development should exploit its natural advantages and include significant, but low-cost improvement in imaging resolution and speed, investigation of clinical benefits of enhanced $T_1$ (and, possibly, $T_2$) contrast, and design of whole-head MEG/ULF-MRI systems. Our work demonstrates that multichannel SQUID systems combining MEG and ULF MRI capabilities for advanced brain studies are practical and efficient. Information provided by such instruments can be easily integrated with data from other imaging modalities, including fMRI, to enable high-resolution spatiotemporal imaging of brain function.

## Acknowledgments

The authors gratefully acknowledge the support of the U.S. National Institutes of Health (Grant No. R01-EB006456), and the U.S. Department of Energy OBER (Grant No. KP150302, project ERWS115). We also thank Dr. Diana South of Mind Research Network (Albuquerque, NM) for performing 1.5 T and 3 T brain scans of our human subject.